\documentclass[a4paper, prl]{revtex4}

\usepackage{amssymb}
\usepackage{amsmath}
\usepackage{graphicx}
\usepackage{psfrag}

\def\slash{\rlap{/}}

\newcommand{\be}{\begin{equation}} \newcommand{\ee}{\end{equation}}
\newcommand{\ba}{\begin{eqnarray}} \newcommand{\ea}{\end{eqnarray}}
\newcommand{\bea}{\begin{eqnarray}} \newcommand{\eea}{\end{eqnarray}}
\newcommand{\bean}{\begin{eqnarray*}} \newcommand{\eean}{\end{eqnarray*}}

\newcommand{\st}{{\scriptscriptstyle T}}

\DeclareMathOperator{\tr}{Tr}

\begin{document}

\title{Gauge Link Structure in Quark-Quark Correlators in hard processes}

\author{C.J. Bomhof}
\email{cbomhof@nat.vu.nl}
\affiliation{
Department of Physics and Astronomy, Vrije Universiteit Amsterdam,\\
NL-1081 HV Amsterdam, the Netherlands}

\author{P.J. Mulders}
\email{pjg.mulders@few.vu.nl}
\affiliation{
Department of Physics and Astronomy, Vrije Universiteit Amsterdam,\\ 
NL-1081 HV Amsterdam, the Netherlands}
                                                                                
\author{F. Pijlman}
\email{f.pijlman@few.vu.nl}
\affiliation{
Department of Physics and Astronomy, Vrije Universiteit Amsterdam,\\
NL-1081 HV Amsterdam, the Netherlands}


\begin{abstract}

Distribution functions in hard processes can be described by 
quark-quark correlators, nonlocal matrix elements of quark fields. 
Color gauge invariance requires inclusion of appropriate gauge 
links in these correlators.
For transverse momentum dependent distribution functions, in particular
important for describing T-odd effects in hard processes, we
find that new link structures containing loops can appear in abelian 
and non-abelian theories. 
In transverse moments, e.g. measured in azimuthal asymmetries, these 
loops may enhance the contribution of gluonic poles. 
Some explicit results for the link structure are given
in high-energy leptoproduction and hadron-hadron scattering.

\end{abstract}

\maketitle

\section{Introduction}
In this paper we discuss the issue of color gauge invariance in
bilocal operator matrix elements off the lightcone~\cite{Ralston:1979ys}.
Such matrix elements are relevant in hard processes in which also transverse
momenta of partons play a role such as semi-inclusive deep inelastic
scattering (SIDIS) or the Drell-Yan process (DY).
In these two cases the quark correlation functions that appear
at leading order in an expansion in the inverse hard scale
turn out to have a different gauge link structure.
Upon integration over transverse momenta the difference in the
gauge link structure does not matter, but for the transverse
moments it does~\cite{Boer:1999si, Belitsky:2002sm, Collins:2002kn, Collins:2003fm}.
For the lowest transverse moments
obtained from transverse momentum dependent distribution functions
the difference corresponds to a time-reversal odd gluonic pole
matrix element~\cite{Boer:2003cm}.
                                                                                
In the tree-level contributions to SIDIS and DY one was in essence
dealing with diagrams with a single `active' quark. 
In this paper we consider more general situations, 
but in order to study the basic features we simplify the discussion by first considering QED-like hard interactions between the quarks. 
Complications that arise in QCD by the explicit presence of gluons will be discussed at the end, 
but the full treatment will be done as part of yet to come applications to hard QCD processes.

\section{Gauge link structures in hard processes}

\begin{figure}[b]
(a)\hspace{1cm}\includegraphics[width=5cm]{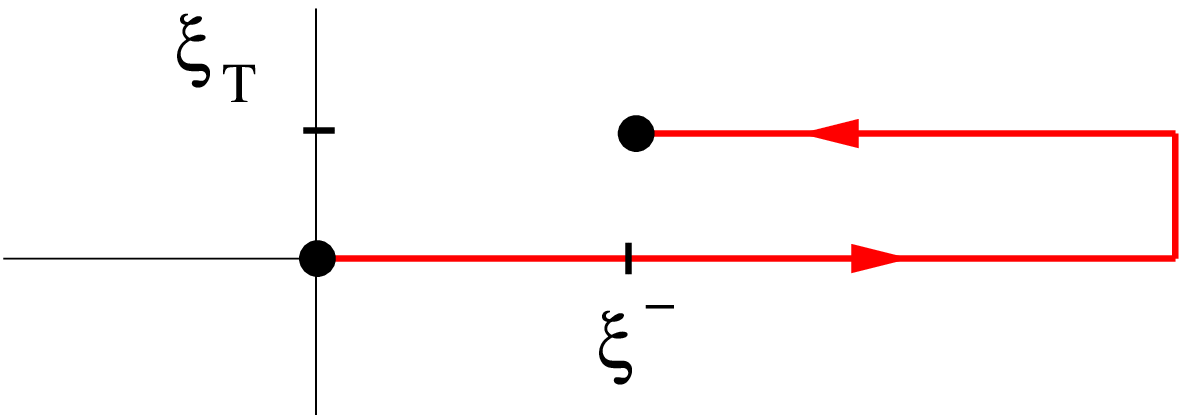}
\hspace{2cm}
(b)\hspace{1cm}\includegraphics[width=5cm]{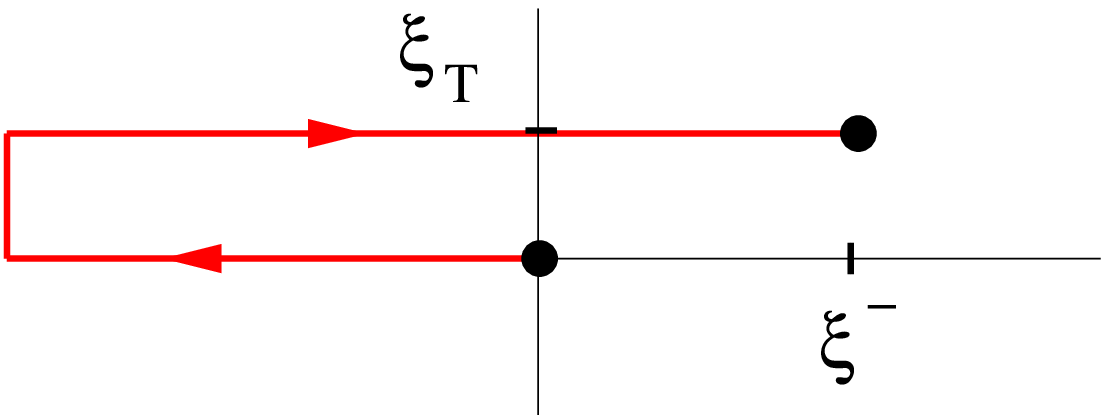}
\caption{The gauge link structure in the quark-quark
correlator $\Phi$ in SIDIS (a) and DY (b) respectively
\label{simplelinks}}
\end{figure}
In electroweak hard scattering processes such as deep inelastic
scattering and the Drell-Yan process, the underlying hard processes
$\gamma^\ast + q \rightarrow q$ and 
$\overline{q} + q \rightarrow \gamma^\ast$
involve a single quark. In these processes a hard scale $Q$ is set
by the virtuality of the photon.
The transition hadron $\rightarrow$ quark
is described by the correlator
\begin{equation}
\Phi_{ij}(p,P) = \int \frac{\mathrm{d}^4 \xi}{(2\pi)^4} \,e^{ip\cdot \xi}
\langle P | \overline{\psi}_j(0) \psi_i (\xi) | P \rangle,
\label{cor-1}
\end{equation}
where depending on the process the nonlocality is limited.
Letting the soft quark and hadron momenta determine the lightcone direction $n_+$, 
one integrates over the quark momentum components $p^- = p\cdot n_+$ 
and $p_\st$ in inclusive deep inelastic scattering (DIS), 
restricting the nonlocality to the 
lightcone~\cite{Collins:1982uw, Soper:1977jc, Jaffe:1983hp}. 
In the case that more external momenta are measured,
implying that other directions can be observed,
one integrates only over the quark momentum component $p^-$, 
restricting the nonlocality to the 
light-front~\cite{Levelt:1994ac, Tangerman:1995eh, Diehl:1998sm}.
Examples are 1-particle inclusive or semi-inclusive deep inelastic
scattering (SIDIS) and the Drell-Yan (DY) process, both of which involve 
two hadrons. 
In that case one has to deal with transverse momentum dependent 
distribution functions.
Lightlike vectors $n_+$ and $n_-$ are introduced for convenience. 
They satisfy $n_- \cdot n_+ = 1$ and are set by the external momenta. 
They define lightcone momenta $a^\pm = a\cdot n_\mp$
and the transverse projector 
$g_\st^{\mu\nu} = g^{\mu\nu}- n_+^{\{\mu}n_-^{\nu\}}$.
Besides the correlator describing the hadron $\rightarrow$ quark transition,
correlators $\overline \Phi(p,P)$ for the hadron 
$\rightarrow$ antiquark and correlators $\Delta(p,P)$ or 
$\overline\Delta(p,P)$ for quark or antiquark $\rightarrow$ hadron transitions
can be written down~\cite{Collins:1982uw,Soper:1977jc}.

Diagrams with additional gluons emerging from the soft hadronic parts
are described by quark-gluon correlators 
\begin{equation}
{\Phi^\alpha_A}_{ij}(p, p-p_1,P) =
\int \frac{\mathrm{d}^4 \xi}{(2\pi)^4} \frac{\mathrm{d}^4 \eta}{(2\pi)^4}
\,e^{ip\cdot \xi} \,e^{ip_1\cdot (\eta-\xi)}
\langle P | \overline{\psi}_j(0) A^\alpha(\eta) \psi_i (\xi) | P \rangle\ .
\end{equation}
In leading order of the expansion in inverse powers of the hard
scale $Q$,
one needs to include the $\Phi^+_A$ correlators involving $A^+ = A\cdot n_-$
(longitudinal) gluons that are collinear with their parent hadron.
Together with multi-gluon correlators they produce gauge links along the $n_-$ direction,
connecting the points $0$ and $\xi$ to lightcone infinity~\cite{Efremov:1981xm}.
Of the correlators involving transverse gluon fields, 
leading parts emerge that involve gluon fields at lightcone $\pm \infty$,
which in combination with the links along $n_-$ complete the gauge links between the points $0$ and $\xi$ 
(see Fig.~\ref{simplelinks})~\cite{Belitsky:2002sm}. 
The resulting gauge links can be included in the correlator in Eq.~\eqref{cor-1},
making it explicitly gauge invariant.
The link structure, 
arising in the off-collinear situation in which one does 
not integrate over all transverse momenta of the quarks,
turns out to have a different path structure for SIDIS as compared to Drell-Yan. 
In the case of SIDIS they come from $A^+$ gluons coupling to an outgoing quark,
while in DY they come from $A^+$ gluons coupling to an incoming antiquark.
The resulting path structures of the link operators are shown in 
Fig.~\ref{simplelinks}.

In this paper we present, in leading order of the inverse hard scale, 
the calculation 
of the gauge link structure in more complex hard processes. 
The hard subprocesses that we will consider in this section are quark-quark
scattering and antiquark-quark scattering,
which we will take as colorless interactions for the sake of simplicity. 
This suffices to illustrate the appearance of additional new structures
in the gauge links.

\begin{figure}
	\begin{center}
		\psfrag{a}[cc][cc]{${\overline\Phi(l)}$}
		\psfrag{b}[cc][cc]{${\Phi(p)}$}
		\psfrag{c}[cc][cc]{${\overline\Delta(l^\prime)}$}
		\psfrag{d}[cc][cc]{${\Delta(k)}$}
		\psfrag{e}[cc][cc]{${M}$}
		\psfrag{f}[cc][cc]{${M^\ast}$}
		\psfrag{g}[cc][cc]{${l}$}
		\psfrag{h}[cc][cc]{${p}$}
		\psfrag{i}[cc][cc]{${l^\prime}$}
		\psfrag{j}[cc][cc]{${k}$}
		\includegraphics[width=6cm]{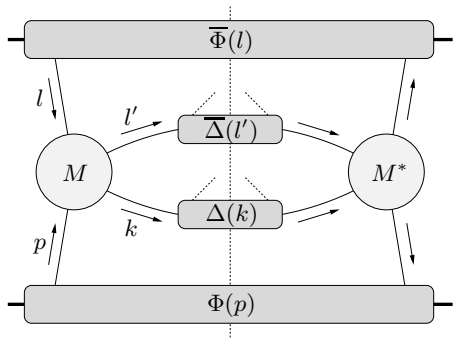}
	\end{center}
	\caption{Decomposition of the cross section for two-hadron production
	in hadron-hadron scattering with a hard antiquark-quark subprocess.
	\label{CROSSSECTION}}
\end{figure}
                                                                               
\begin{figure}[t]
(a)\hspace{0.5cm}\includegraphics[width=7.0cm]{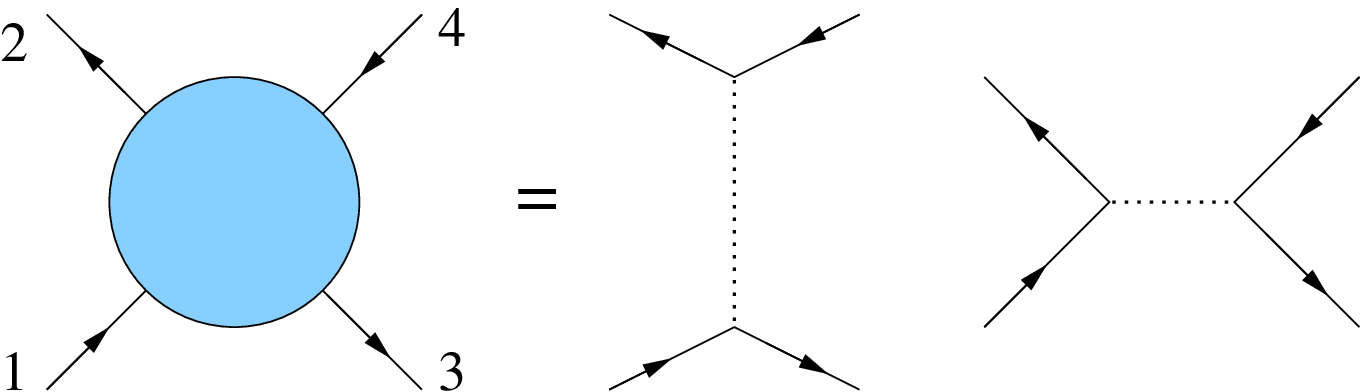}
\hspace{1cm}
(b)\hspace{0.5cm}\includegraphics[width=7.0cm]{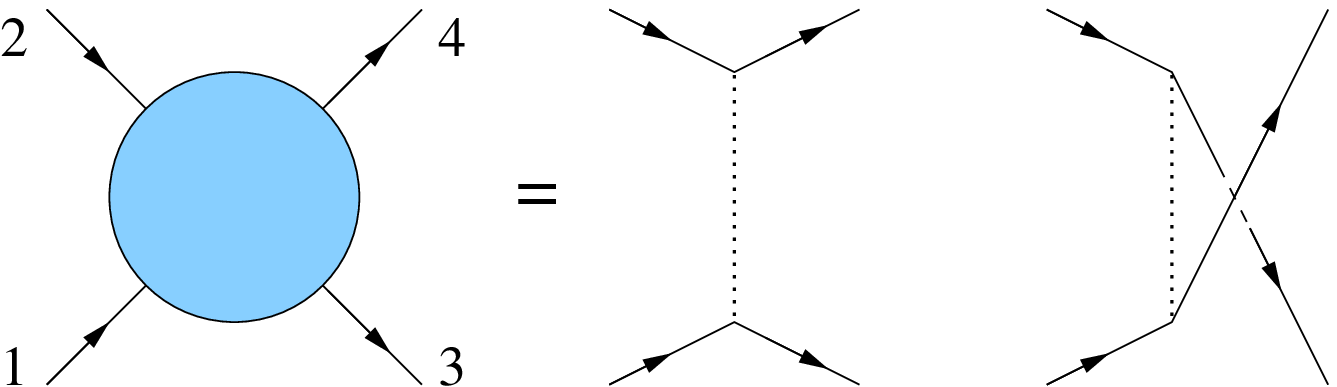}
\caption{
(a) The t- and s-channel contributions to quark-antiquark scattering 
($q_1 + \overline q_2 \rightarrow q_3 + \overline q_4$). 
(b) The t- and u-channel contributions to 
quark-quark scattering ($q_1 + q_2 \rightarrow q_3 + q_4$).
\label{Processes}}
\end{figure}

The starting point in the calculation of high-energy hadronic scattering 
cross sections is the assumption that at tree-level it can be written as a convolution of hard, 
perturbatively calculable, partonic scattering processes with
(the aforementioned) correlation functions describing the 
hadron $\rightarrow$ quark or quark $\rightarrow$ hadron 
transitions~\cite{Politzer:1980me, Ellis:1982wd, Ellis:1983cd}.
For example, the tree-level contribution to the
cross section of the high-energy hadronic scattering process
$h_A+h_B\longrightarrow h_C+h_D+X$ (see Fig.~\ref{CROSSSECTION}) involving
the amplitude $M$ for the hard quark-antiquark scattering
process $q_1(p) + \overline q_2(l)
\longrightarrow q_3(k) + \overline q_4(l^\prime)$ is written as
\begin{equation}
\sigma \quad\propto\quad 
\int \mathrm{d}^4 p\,\mathrm{d}^4 k\,\mathrm{d}^4 l\,\mathrm{d}^4 l^\prime 
\ \Phi(p)\otimes\overline\Phi(l)\otimes\vert M(p,k,l,l^\prime)\vert^2
\otimes\Delta(k)\otimes\overline\Delta(l^\prime).
\label{GeneralSigma}
\end{equation}
The hard partonic scattering may contain several contributions, 
see Fig.~\ref{Processes}a. 
The correlators $\Phi$, $\overline \Phi$, 
$\Delta$ and $\overline \Delta$ describe the soft parts connecting the quarks in the hard process to the 
hadrons involved in the physical (partly inclusive) process. 
The convolution  in equation~\eqref{GeneralSigma} indicates that one needs specific Dirac tracing, 
depending on the particular forms of $M$. 

\begin{figure}[b]
	\psfrag{b}[rc][rc]{${l}$}
	\psfrag{c}[cb][cb]{${k}$}
	\psfrag{d}[cb][cb]{${l^\prime}$}
	\psfrag{f}[lc][lc]{${p_1}$}
	\centering
	\begin{minipage}[t]{6cm}
		\centering
		\psfrag{a}[rc][rc]{${p}$}
		\psfrag{e}[rc][rc]{${q}$}
		\includegraphics[width=6cm]{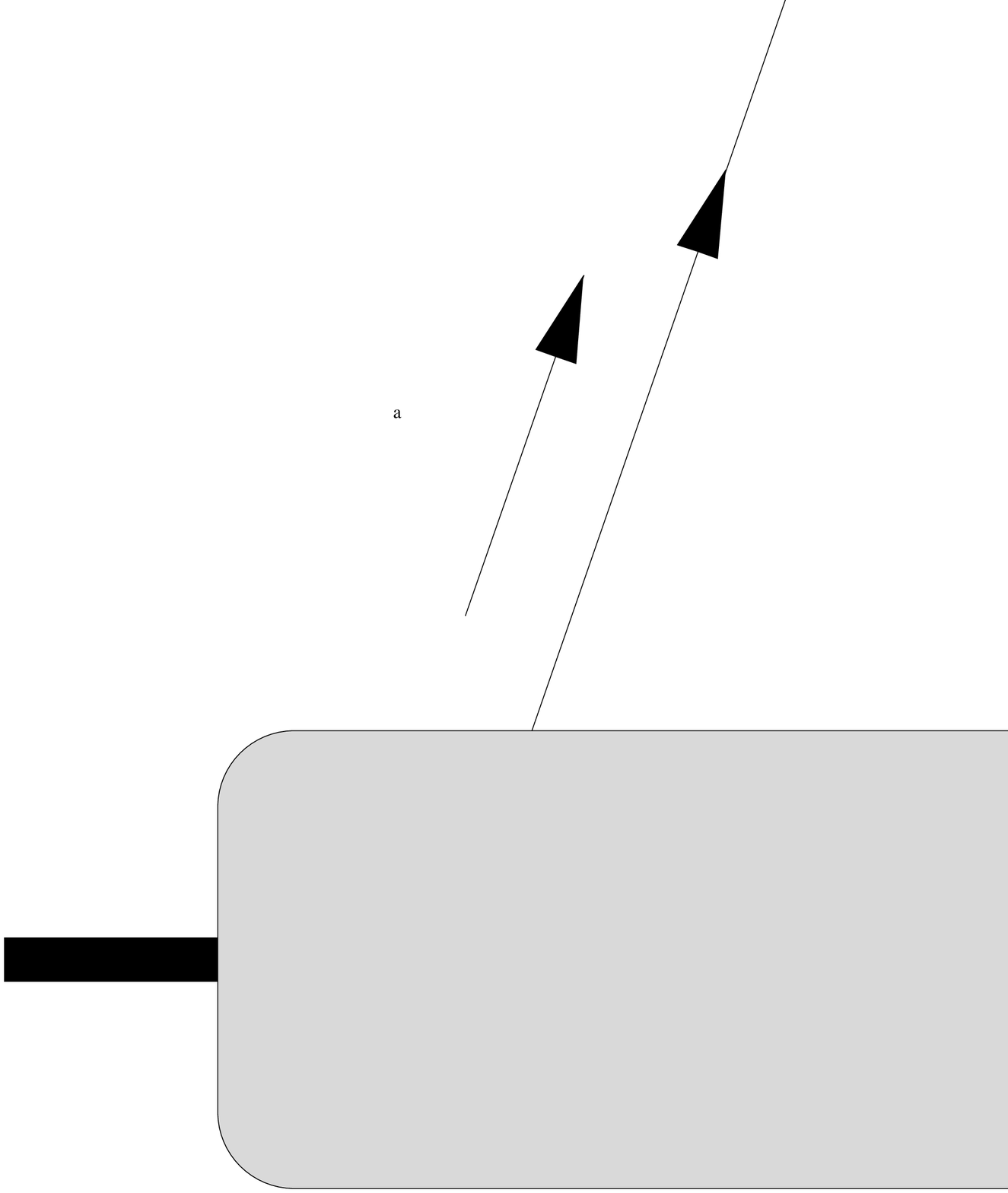}\\
		a
	\end{minipage}\hspace{10mm}%
	\begin{minipage}[t]{6cm}
		\centering
		\psfrag{a}[rc][rc]{${p-p_1}$}
		\psfrag{e}[rc][rc]{${q}$}
		\includegraphics[width=6cm]{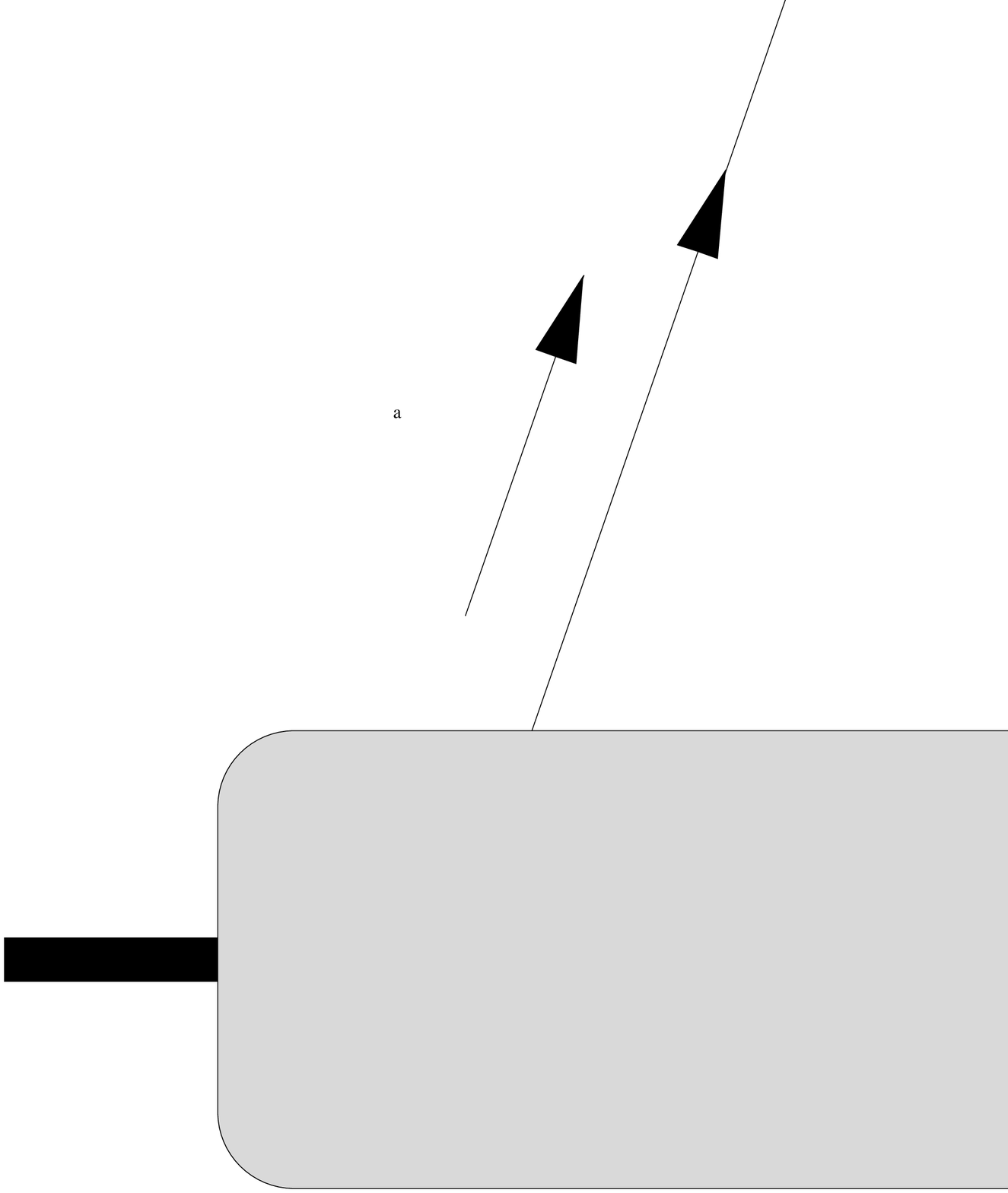}\\
		b
	\end{minipage}\\[4mm]
	\begin{minipage}[t]{6cm}
		\centering
		\psfrag{a}[rc][rc]{${p-p_1}$}
		\psfrag{e}[rc][rc]{${q+p_1}$}
		\includegraphics[width=6cm]{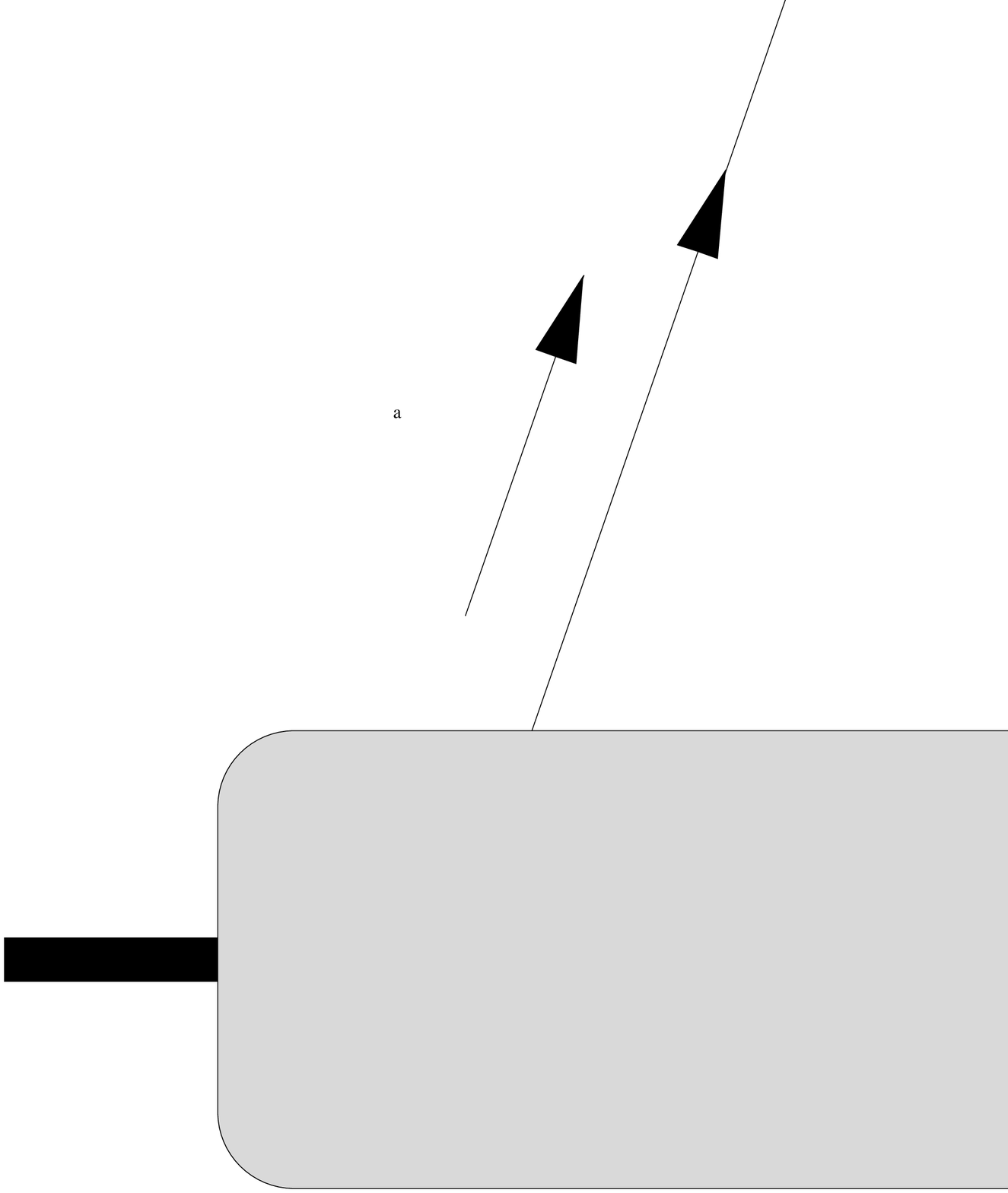}\\
		c
	\end{minipage}\hspace{10mm}%
	\begin{minipage}[t]{6cm}
		\centering
		\psfrag{a}[rc][rc]{${p-p_1}$}
		\psfrag{e}[rc][rc]{${q+p_1}$}
		\includegraphics[width=6cm]{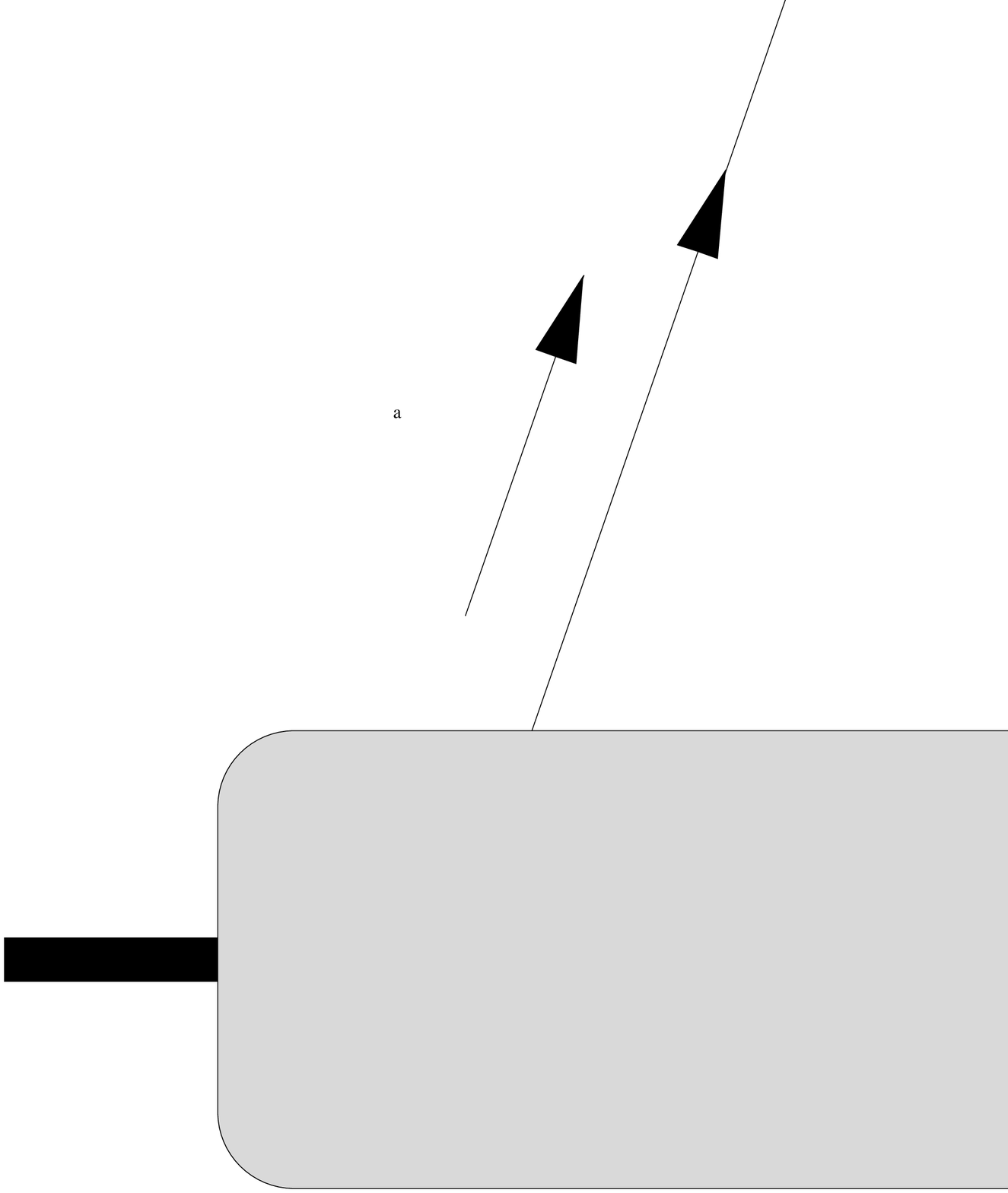}\\
		d
	\end{minipage}
	\caption{Treelevel diagram and insertions of longitudinal gauge bosons.
		\label{Fig1}}
\end{figure}

As an illustration of equation~\eqref{GeneralSigma}, 
let us consider the situation where the incoming quark and antiquark have different flavors, 
in which case one only has the t-channel contribution in the hard process in Fig.~\ref{Processes}a. 
Because we want to illustrate the involvement of another strongly interacting particle in the hard process, 
it is convenient to keep the hard process itself colorless.
Therefore, we consider the exchange of a photon as indicated in Figure~\ref{Fig1}a.
The expression for the scattering cross section is
\begin{equation}
\sigma\quad\propto\quad
\int \mathrm{d}^4 p\,\mathrm{d}^4 k\,\mathrm{d}^4 l\,\mathrm{d}^4 l^\prime
\ \delta^4(p+q-k) 
\bigg\{\,\Big(\frac{1}{q^2}\Big)^2\,
\tr\big[\Phi(p)\gamma^\mu\Delta(k)\gamma^\nu\big]\,
\tr\big[\,\overline\Phi(l)\gamma_\nu\overline\Delta(l^\prime)\gamma_\mu\big]\,\bigg\}
\end{equation}
where $q=l'-l$.
To obtain the link structure of the $\Phi(p)$ correlator in equation~\eqref{GeneralSigma},
we consider all the (colorless) gluon insertions coming from the lower blob.
All other soft correlators in the hadronic scattering 
process also get specific gauge link structures from their respective
longitudinal gluons coupling to the hard part.
Introducing a set of light-like vectors $n_+$ and $n_-$, 
such that $n_+$ is proportional to the momentum of the lower blob
and $n_-\cdot n_+=1$,
we find that the single $A^+=A\cdot n_-$ gluon insertions
on the l.h.s.\ of the cut lead to the additional contributions  
in the cross-section in Figs.~\ref{Fig1}b-d
\begin{eqnarray}
\sigma 
&\propto&
\int \mathrm{d}^4 p\,\mathrm{d}^4 k\,\mathrm{d}^4 l\,\mathrm{d}^4 l^\prime\ 
\delta^4(p+q-k)\ \frac{1}{q^2}\ 
\bigg\{\frac{1}{q^2+i\epsilon}\,
\tr\big[\Phi(p)\gamma^\mu\Delta(k)\gamma^\nu\big]\,
\tr\big[\,\overline\Phi(l)\gamma_\nu\overline\Delta(l^\prime)\gamma_\mu\big]
\nonumber \\
&&\mbox{}+ g_2\int \mathrm{d}^4 p_1\ \frac{1}{q^2+i\epsilon}\,
\tr\Big[\Phi_A^\alpha(p,p-p_1)\gamma^\mu\Delta(k)(i\gamma_\alpha)
\frac{i(\slash{k}-\slash{p_1})}{(k-p_1)^2 + i \epsilon} \gamma^\nu\Big]
\tr\big[\,\overline\Phi(l)\gamma_\nu\overline\Delta(l^\prime)\gamma_\mu\big] 
\nonumber \\
&&\mbox{} + g_1\int \mathrm{d}^4 p_1\ \frac{1}{(q+p_1)^2+i\epsilon}\,
\tr\big[\Phi_A^\alpha(p,p-p_1)\gamma^\mu\Delta(k)\gamma^\nu\big] 
\tr\Big[\overline\Phi(l)\gamma_\nu
\frac{i(-\slash{l}^{\ '}+\slash{p_1})}{(l'-p_1)^2+i\epsilon}(i\gamma_\alpha)
\overline\Delta(l^\prime)\gamma_\mu\Big]
\nonumber \\
&&\mbox{}+g_1\int \mathrm{d}^4 p_1\ \frac{1}{(q+p_1)^2+i\epsilon}\,
\tr\big[\Phi_A^\alpha(p,p-p_1)\gamma^\mu\Delta(k)\gamma^\nu\big] 
\tr\Big[\overline\Phi(l)(i\gamma_\alpha)
\frac{i(-\slash{l}-\slash{p_1})}{(l+p_1)^2 + i \epsilon}
\gamma_\nu\overline\Delta(l^\prime)\gamma_\mu\Big]\,\bigg\}\label{INSERTIONS}
\end{eqnarray}
where $p$ and $p_1$ are collinear with $n_+$. 
For ease of distinguishing the involved fermions, 
the coupling constants of the longitudinal gauge particles are indicated
with $g_1$ and $g_2$. 
Investigating the analytic structure in $p_1^+$, 
one finds that the poles at $p_1^+ \neq 0$ cancel each other. 
This is analogous to Ward identities,
where consecutive insertions take care of all the cancellations of the poles
at $p_1^+ \neq 0$.
Making use of the fact that for the leading contributions in the soft
correlators $\Phi(p) \propto \slash p$, $\Delta (k) \propto \slash k$,
$\overline \Phi(l) \propto \slash l$ and $\Delta (l^\prime) \propto
\slash l^{\ \prime}$~\cite{Barone:2001sp},
the final result for the scattering cross section is 
\begin{eqnarray}
\sigma&\propto&
\int \mathrm{d}^4 p\,\mathrm{d}^4 k\,\mathrm{d}^4 l\,\mathrm{d}^4 l^\prime\ 
\delta^4(p+q-k)\ \Big(\frac{1}{q^2}\Big)^2\ \bigg\{
\tr\big[\Phi(p)\gamma^\mu\Delta(k)\gamma^\nu\big]
\tr\big[\,\overline\Phi(l)\gamma_\nu\overline\Delta(l^\prime)\gamma_\mu\big] \nonumber \\
&&\ -g_2\int \mathrm{d}^4 p_1 \frac{1}{-p_1^+ + i \epsilon}
\tr\big[\Phi_A^\alpha(p,p-p_1) \gamma^\mu\Delta(k)\gamma^\nu\big]
\tr\big[\,\overline\Phi(l)\gamma_\nu\overline\Delta(l^\prime)\gamma_\mu\big]
\nonumber \\
&&\ +g_1\int \mathrm{d}^4 p_1 \frac{1}{-p_1^+ + i \epsilon}
\tr\big[\Phi_A^\alpha(p,p-p_1) \gamma^\mu\Delta(k)\gamma^\nu\big]
\tr\big[\,\overline\Phi(l)\gamma_\nu\overline\Delta(l^\prime)\gamma_\mu\big]
\nonumber \\
&&\ +g_1\int \mathrm{d}^4 p_1 \frac{1}{p_1^+ + i \epsilon}
\tr\big[\Phi_A^\alpha(p,p-p_1) \gamma^\mu\Delta(k)\gamma^\nu\big]
\tr\big[\,\overline\Phi(l)\gamma_\nu\overline\Delta(l^\prime)\gamma_\mu\big]\,
\bigg\}
\label{linkfirstorder}
\end{eqnarray}

\begin{figure}[b]
(a)\hspace{0.5cm}\includegraphics[width=5.5cm]{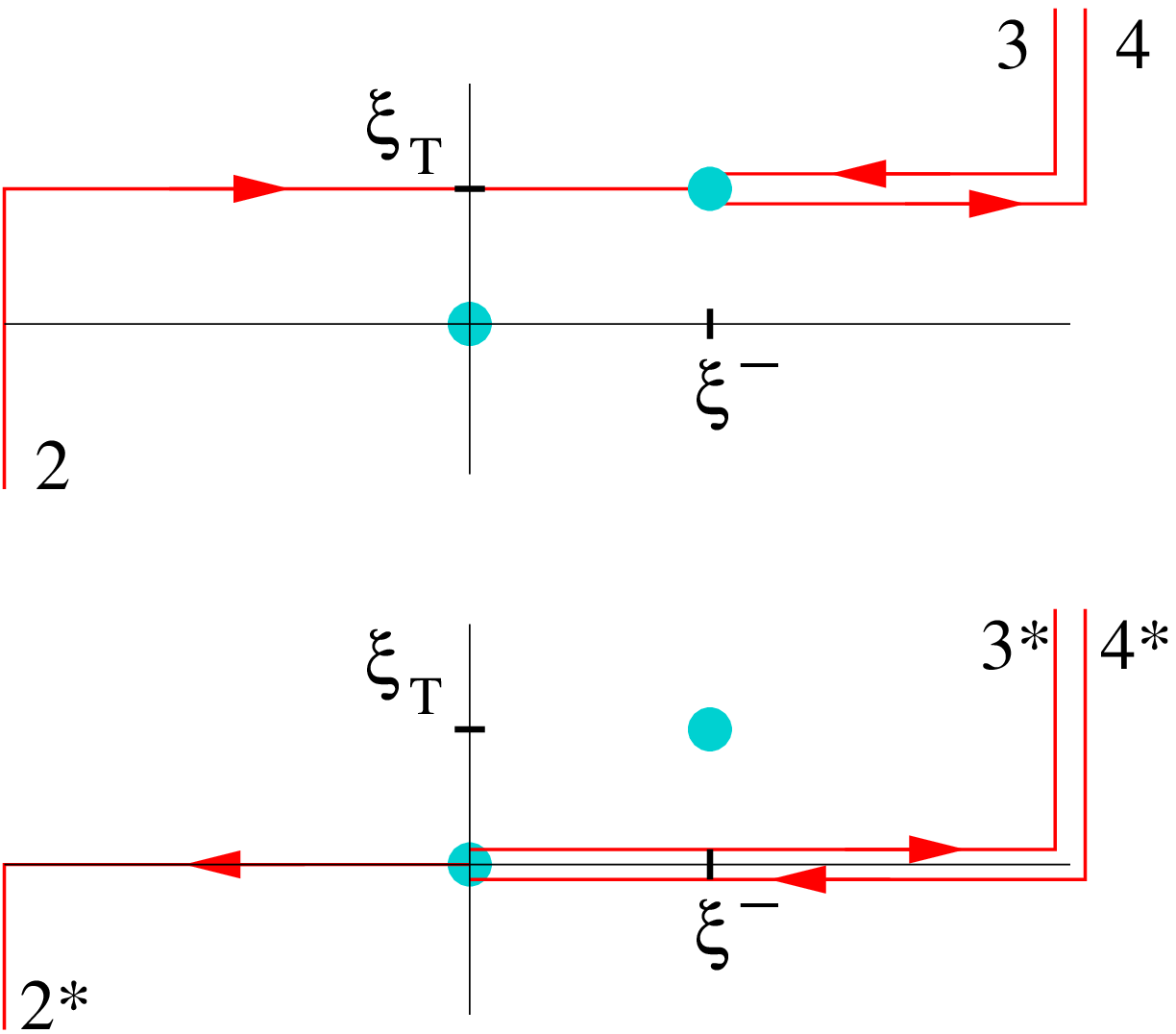}
\hspace{2cm}
(b)\hspace{0.5cm}\includegraphics[width=5.5cm]{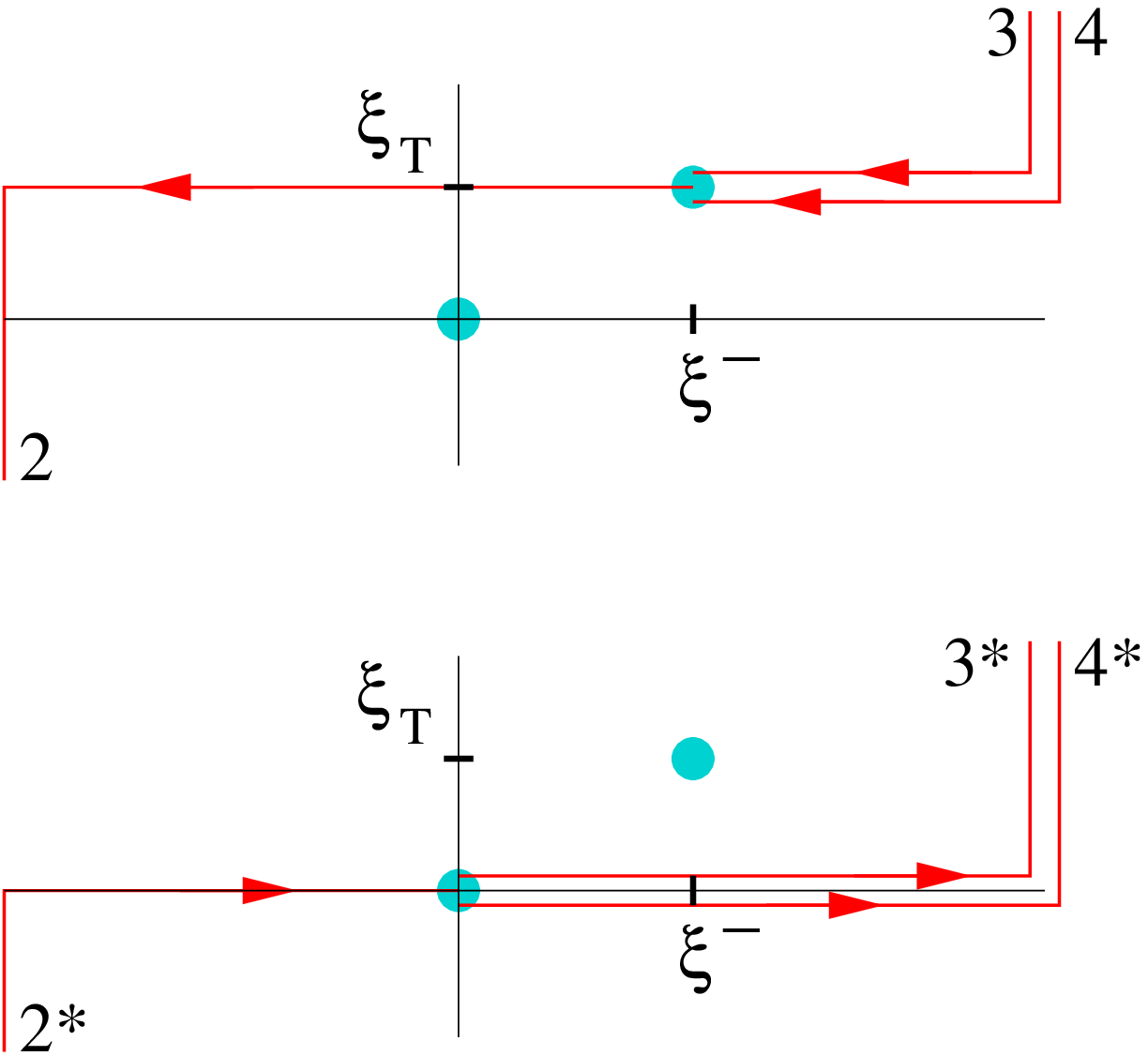}
\caption{The basic gauge links for distribution functions 
coming from the soft correlator for quark 1 that arise from
$A^+$-gluons interacting with the other initial- or final-state
fermion legs in the amplitude (2, 3 and 4) and the conjugate amplitude
(2*, 3* and 4*) for 
(a) quark-antiquark scattering
($q_1 + \overline q_2 \rightarrow q_3 + \overline q_4$) 
(b) quark-quark scattering ($q_1 + q_2 \rightarrow q_3 + q_4$).
\label{fig-processlinks}}
\end{figure}

The contributions found in Eq.~\eqref{linkfirstorder} represent the first terms of gauge links 
$U_{g_2}(\infty^-,\xi_T;\xi^-,\xi_T)$, 
$U_{g_1}(\xi^-,\xi_T;\infty^-,\xi_T)$, and
$U_{g_1}(-\infty^-,\xi_T;\xi^-,\xi_T)$, respectively. 
They are of the form
\begin{equation}
U_g(a;b) = \mathcal{P} \exp \bigg( -i g \int_a^b \mathrm{d}x \cdot A(x)
\bigg),
\end{equation}
with the integration along a straight line between $a$ and $b$. 
The link structure arising in this way from the insertions 
of longitudinal gluons coming from $\Phi(p)$ (Figs~\ref{Fig1}b-d)
in the hard amplitude $q_1 \overline q_2 \rightarrow q_3 \overline q_4$ 
is indicated in Fig.~\ref{fig-processlinks}a. The same figure also
gives the results for the insertions in the conjugate diagram, 
which can be handled in the same way and
lead to the links connecting the point $0$ in the correlation
function to lightcone infinity.
The first order calculations presented here
explicitly, can be extended to include all longitudinal gluon insertions
to which we,  
without giving the derivation, have added the transverse gauge link
pieces.
These 
transverse fields at infinity emerge, 
in the same way as for the simple processes with just a single quark line,
as boundary terms that also need to be substracted from 
transverse gluon correlators $\Phi_A^\alpha$ to obtain gauge
invariant correlators in terms of the field strength tensor~\cite{Boer:2003cm}.
For a hard quark-quark scattering amplitude 
$q_1 q_2 \rightarrow q_3 q_4$ the resulting link structures are given in
Fig.~\ref{fig-processlinks}b.

\begin{figure}[b]
(a)
\includegraphics[width=3cm]{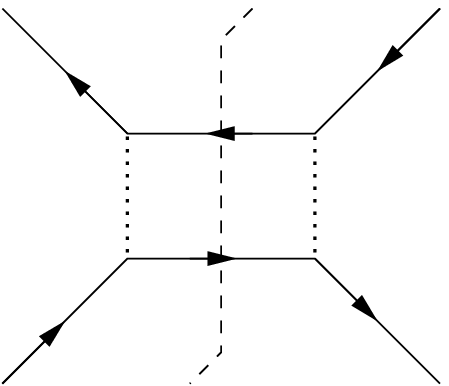}\hspace{2cm}
\includegraphics[width=5.5cm]{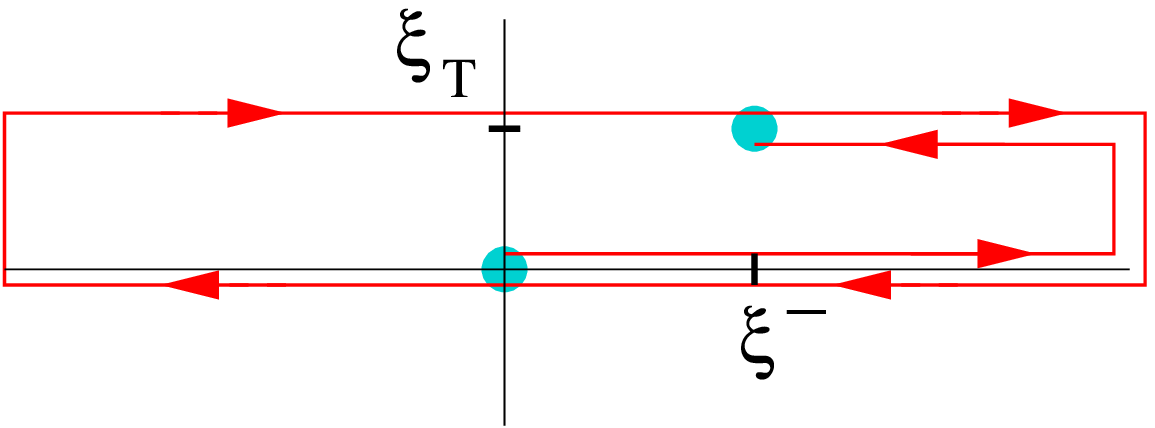}
\hspace{1cm}
\begin{minipage}{3cm}
\[
\mathcal{U}^{[+]}_{g_2}
\ \tr\big(\mathcal{U}^{[\Box]\dagger}_{g_1}\big)
\]
\\[1.5cm]\mbox{}
\end{minipage}
\\[-1.0cm]    
(b)\includegraphics[width=3cm]{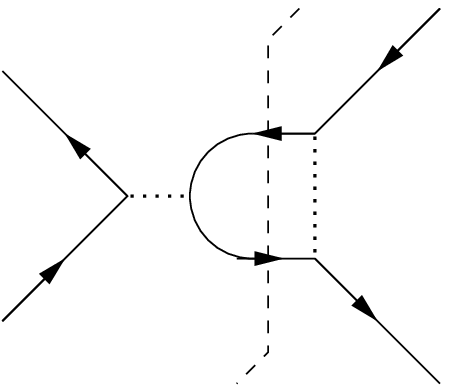}\hspace{2cm}
\includegraphics[width=5.5cm]{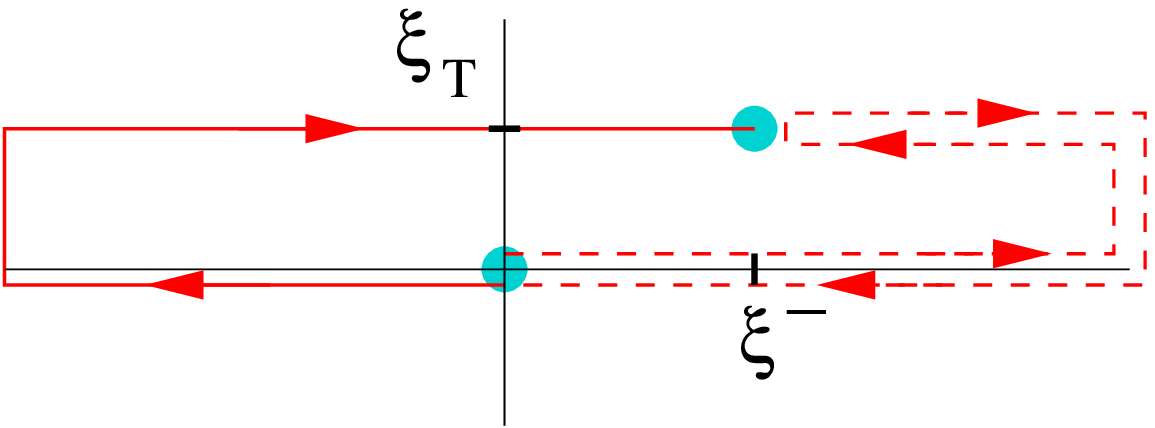}\hspace{1cm}
\begin{minipage}{3cm}
\[
\mathcal{U}^{[-]}_{g}
\]
\\[1.5cm]\mbox{}
\end{minipage}
\\[-1.0cm]    
(c)\includegraphics[width=3cm]{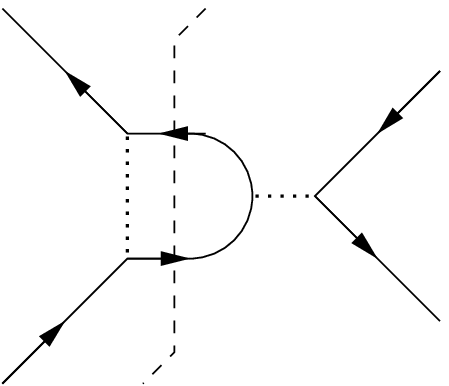}\hspace{2cm}
\includegraphics[width=5.5cm]{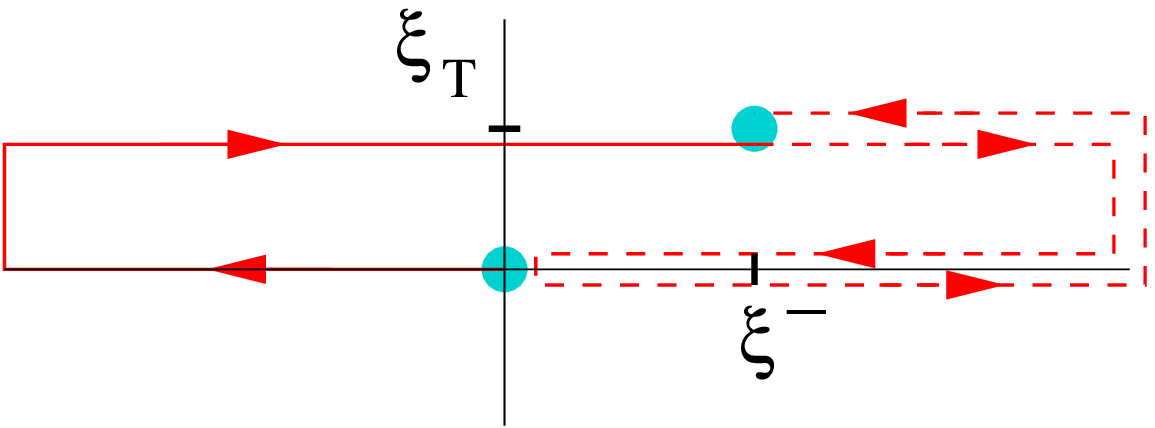}\hspace{1cm}
\begin{minipage}{3cm}
\[
\mathcal{U}^{[-]}_{g}
\]
\\[1.5cm]\mbox{}
\end{minipage}
\\[-1.0cm]    
(d)\includegraphics[width=3cm]{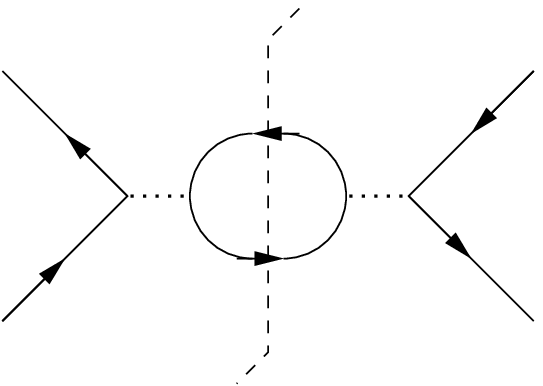}\hspace{2cm}
\includegraphics[width=5.5cm]{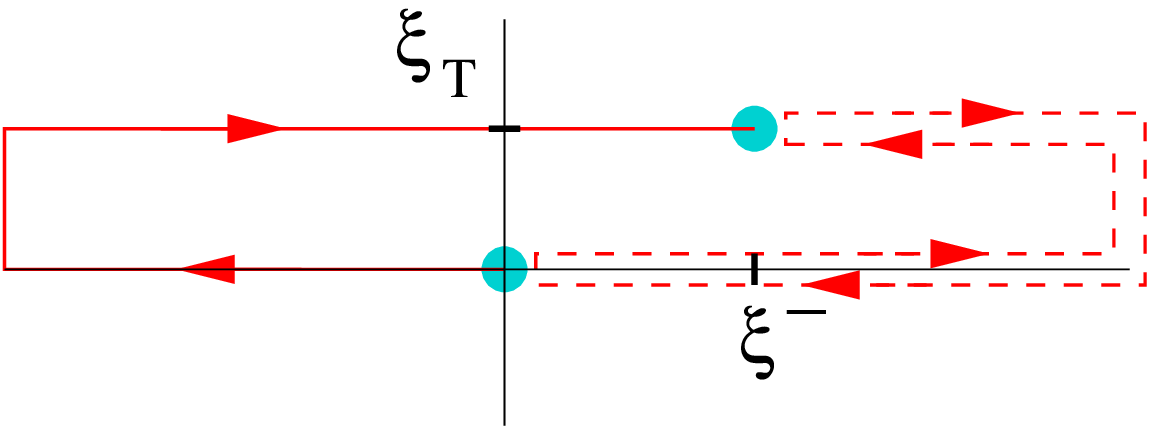}\hspace{1cm}
\begin{minipage}{3cm}
\[
\mathcal{U}^{[-]}_{g}
\ \tr\big(\mathbb I\big)
\]
\\[1.5cm]\mbox{}
\end{minipage}
\\[-1.0cm]    
\caption{Contributions to the cross section of antiquark-quark
scattering and the corresponding gauge link structures. 
In the contributions b-d one has $g_1 = g_2=g$.
\label{fig-qqbarlinks}}
\end{figure}

Taking all the insertions of longitudinal gluons into account,
as in equation~\eqref{INSERTIONS}, 
we have found that this expression reduces to equation~\eqref{GeneralSigma} with the correlator $\Phi(p)$ now containing a link connecting the quark fields at $0$ and $\xi$.
Three combinations of links appear
\begin{eqnarray}
\mathcal{U}^{[-]}_g 
&=& U_g(0;-\infty^-,0_T) U_g(-\infty^-,0_T; -\infty^-,\xi_T) U_g(\infty^-,\xi_T;\xi)\ ,\label{loop-dy}\\
\mathcal{U}^{[+]}_g 
&=& U_g(0;\infty^-,0_T) U_g(\infty^-,0_T; \infty^-,\xi_T) U_g(\infty^-,\xi_T;\xi)\ ,\label{loop-dis}\\
\mathcal{U}^{[\Box]}_g 
&=&
\mathcal{U}^{[+]}_g\,\mathcal{U}^{[-]\dagger}_g\ ,
\end{eqnarray}
where the latter constitutes a counterclockwise (Wilson) loop from the point
$0$ via lightcone infinity to $\xi$ and back via negative lightcone
infinity to $0$.
The result for $h_A + h_B \rightarrow h_C + h_D+X$ with the quark-antiquark
subprocess is a gauge link insertion 
$\mathcal{U}^{[+]}_{g_2}\,\tr (\mathcal{U}^{[\Box]\dagger}_{g_1})$,
shown in Fig.~\ref{fig-qqbarlinks}a. 
The trace operation introduced in this figure will become relevant when we take the full color structure of the inserted gluons into account.

An interesting case to consider is when $q_1$ and $\overline q_2$ are each others antiparticles.
Besides the t-channel contribution which we have treated above, 
there is also the s-channel quark-antiquark annihilation contribution
(see Fig.~\ref{Processes}a). 
The link structure for the s-channel amplitude and its conjugate are, 
just as the t-channel contribution,
given by the gauge lines in Fig.~\ref{fig-processlinks}a. 
Including interference terms,
four contributions arise in the cross section. 
We observe that in these terms the correlation function $\Phi(p)$ appears 
with different gauge link structures as indicated in 
Fig.~\ref{fig-qqbarlinks}b-d. 
In the terms of the scattering cross section that contain an s-channel 
amplitude, we again get the product of the link operators
$\mathcal{U}^{[+]}_{g_2}$ and $\mathcal{U}^{[\Box]\dagger}_{g_1}$,
but in these terms we only have a single coupling constant $g_1=g_2=g$ and 
the two link operators add up to the link structure $\mathcal{U}^{[-]}_{g}$.
We note that the result in Fig.~\ref{fig-qqbarlinks}d actually involves
$\tr\big(\mathcal{U}^{[+]}_{g}\,\mathcal{U}^{[+]\dagger}_{g}\big)$,
i.e. the trace of the unity operator in the charge space, 
which becomes relevant in case the color structure is considered.

\begin{figure}[b]
\begin{center}
(a)\includegraphics[width=2.5cm]{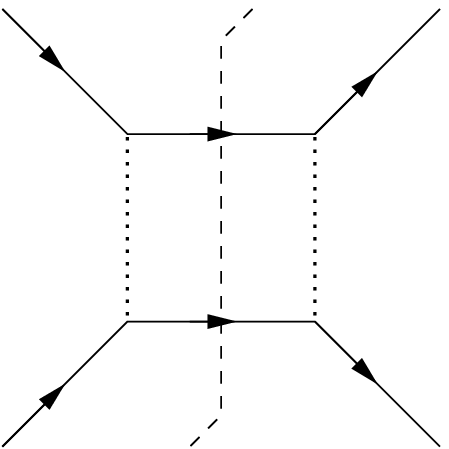}\hspace{2cm}
\includegraphics[width=5.5cm]{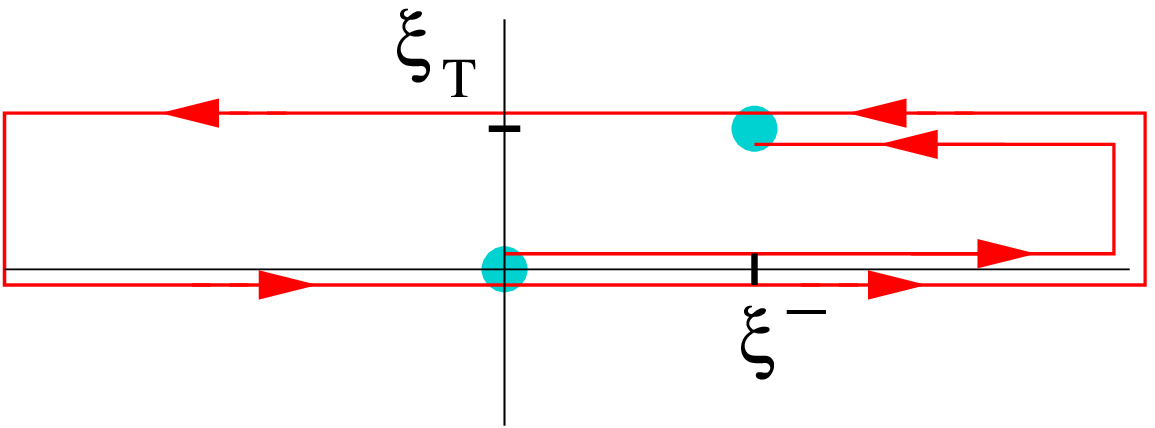}
\hspace{1cm}
\begin{minipage}{3cm}
\[
\tr\big(\mathcal{U}^{[\Box]}_{g_1}\big)
\,\mathcal{U}^{[+]}_{g_2}
\]
\\[1.5cm]\mbox{}
\end{minipage}
\\[-1.0cm]    
(b)\includegraphics[width=2.5cm]{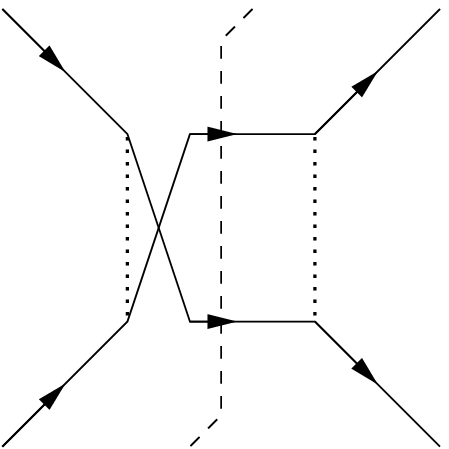}\hspace{2cm}
\includegraphics[width=5.5cm]{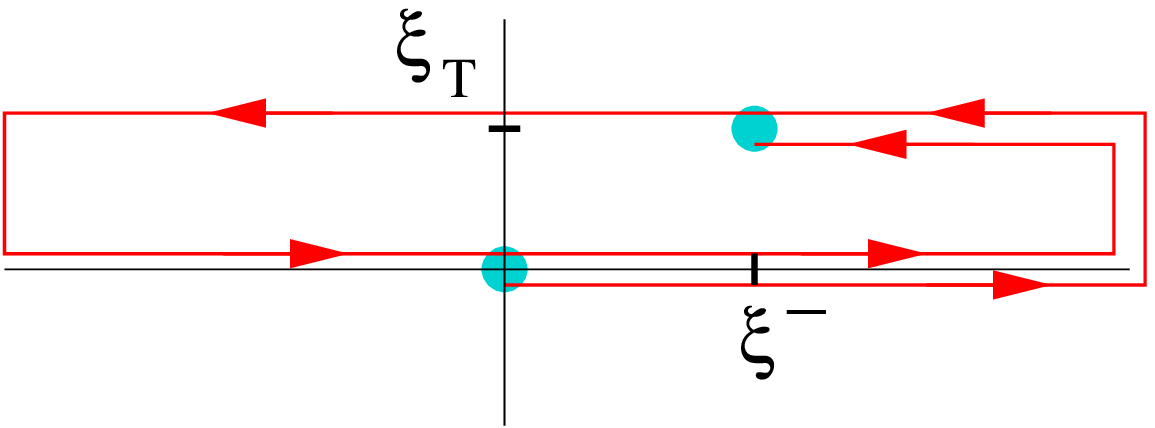}\hspace{1cm}
\begin{minipage}{3cm}
\[
\mathcal{U}^{[\Box]}_g
\,\mathcal{U}^{[+]}_{g}
\]
\\[1.5cm]\mbox{}
\end{minipage}
\\[-1.0cm]    
(c)\includegraphics[width=2.5cm]{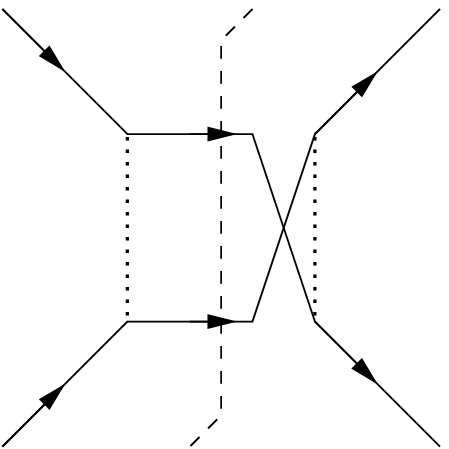}\hspace{2cm}
\includegraphics[width=5.5cm]{Figures/link4b.eps}\hspace{1cm}
\begin{minipage}{3cm}
\[
\mathcal{U}^{[\Box]}_g
\,\mathcal{U}^{[+]}_{g}
\]
\\[1.5cm]\mbox{}
\end{minipage}
\\[-1.0cm]    
(d)\includegraphics[width=2.5cm]{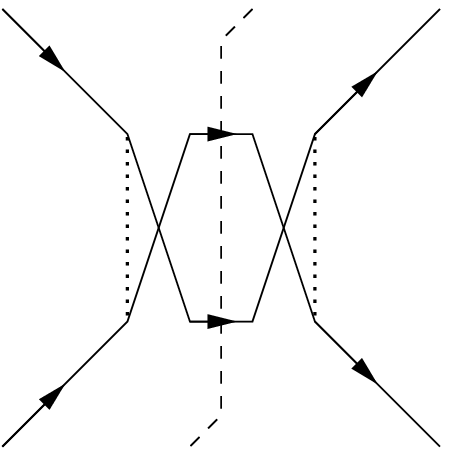}\hspace{2cm}
\includegraphics[width=5.5cm]{Figures/link4a.eps}\hspace{1cm}
\begin{minipage}{3cm}
\[
\tr\big(\mathcal{U}^{[\Box]}_g\big)
\,\mathcal{U}^{[+]}_{g}
\]
\\[1.5cm]\mbox{}
\end{minipage}
\\[-1.0cm]    
\caption{Contributions to the cross section of quark-quark
scattering and the corresponding gauge link structures.
In the contributions b-d one has $g_1 = g_2=g$.
\label{fig-qqlinks}}
\end{center}
\end{figure}

If we take the hard process $M(p,k,l,l')$ in Fig.~\ref{CROSSSECTION} to represent quark-quark scattering, 
the gauge links arising from the insertions of collinear gluons in the amplitude and its conjugate (Fig.~\ref{fig-processlinks}b) 
combine into one or (in case of identical quarks) four possible link structures in the cross section,
as summarized in Fig.~\ref{fig-qqlinks}.
Again we get the product of the link operators
$\mathcal{U}^{[+]}_{g_2}$ and $\mathcal{U}^{[\Box]}_{g_1}$,
where $\mathcal{U}^{[\Box]}_{g_1}$ is traced if the amplitudes on both sides of the cut are each others hermitian conjugates.
For the interference terms in quark-quark scattering, however, 
the two operators do not add up to a simple link operator such as 
$\mathcal{U}^{[-]}_{g}$.
 
The loop that we have found in the gauge link disappears upon integration of the matrix elements over transverse momentum. 
Therefore, it only affects those processes where one is sensitive to the transverse momentum of the quarks inside the nucleon.
These involve transverse moments $\Phi_\partial^\alpha (x)$,
defined as transverse momentum weighted integrals of the $p_\st$-dependent
distribution functions.
In these moments the loop will contribute through a gluonic pole matrix element, responsible for time-reversal odd distribution functions, 
such as the Sivers function~\cite{Sivers:1990cc, Sivers:1991fh}.
For example, 
for unpolarized quarks in SIDIS one has~\cite{Boer:2003cm}
\begin{eqnarray}
&&\tr\big[ \Phi_\partial^{[+]} (x) \gamma^+ \big] 
=\int \mathrm{d}^2 p_\st\ p_\st^\alpha 
\int \frac{\mathrm{d}^2 \xi_\st \mathrm{d} \xi^-}{(2\pi)^3}\ e^{ip\cdot\xi} 
\langle P,S\vert \overline{\psi}(0) \gamma^+\mathcal U_{g}^{[+]}(0;\xi) 
\psi(\xi) \vert P,S \rangle \Big|_{\xi^+=0} \nonumber\\
&&\qquad \mbox{}
= \frac{g}{2} \int \frac{\mathrm{d} \xi^- }{2\pi}\ e^{ip^+\xi^-}  
\ \langle P,S \vert \overline{\psi}(0) \gamma^+ 
\int_{-\infty}^\infty \mathrm{d}\eta^-\ 
\mathcal U(0;\eta^-) G^{+\alpha} (\eta^-)\mathcal U(\eta^-;\xi^-) \psi ( \xi) \vert P,S 
\rangle\Big|_{\xi^+ = \xi_\st = 0}\ ,
\label{glp0}
\end{eqnarray}
(only the T-odd part).
Transverse moments appearing in processes involving hard quark-quark scattering yield different results. 
Taking for instance the link structure in Fig.~\ref{fig-qqlinks}b gives
\begin{eqnarray}
&&
\int \mathrm{d}^2 p_\st\ p_\st^\alpha \int \frac{\mathrm{d}^2 
\xi_\st \mathrm{d} \xi^-}
{(2\pi)^3}\ e^{ip\cdot\xi} \langle P,S\vert \overline{\psi}(0) \gamma^+ 
\mathcal U_{g}^{[\Box]}\,\mathcal U_{g}^{[+]}(0;\xi) \psi(\xi) | 
P,S \rangle \Big|_{\xi^+=0} \nonumber\\
&&\qquad \mbox{} =  
3\,\frac{g}{2}  \int \frac{\mathrm{d} \xi^-}{2\pi}\ e^{ip^+\xi^-}\ 
\langle P,S | \overline{\psi}(0) \gamma^+ 
\int_{-\infty}^\infty \mathrm{d}\eta^-
\ \mathcal U(0;\eta^-) G^{+\alpha} (\eta^-)\mathcal U(\eta^-;\xi^-) 
\psi (\xi) | P,S \rangle\Big|_{\xi^+ = \xi_\st = 0}\ ,
\label{glp}
\end{eqnarray}
where one finds that the difference is again a gluonic pole matrix element,
but with a different strength. 
This strength is set by the link structure and thus by the process, 
a feature that also appears in non-abelian theories,
as we will see in the next section.

%

\section{Link structures in QCD processes}

Although we have made various simplifying assumptions in the processes that we have considered in the previous section, 
it illustrates the appearance of open and closed integration paths in the gauge link in the correlator $\Phi$, 
which we will also find in QCD. 
In the abelian examples given above the only particles carrying charge were the fermions, 
hence the charge tracing followed the Dirac tracing. 
In the non-abelian case such as QCD,
this is in general no longer the case.
In the previous section, 
we already indicated the charge flow and contractions via the traces in
Figs~\ref{fig-qqbarlinks} and \ref{fig-qqlinks},
which is particularly relevant in QCD.
The non-abelian loops are not gauge invariant by themselves, 
but only together with the quark fields in the correlator or when they are color-traced.
In the calculation of several 
QCD processes we find that the sum of all gluon insertions 
to a single amplitude again does not have any poles at $p_1^+ \neq 0$.

\begin{figure}
\begin{center}
\includegraphics[width=4cm]{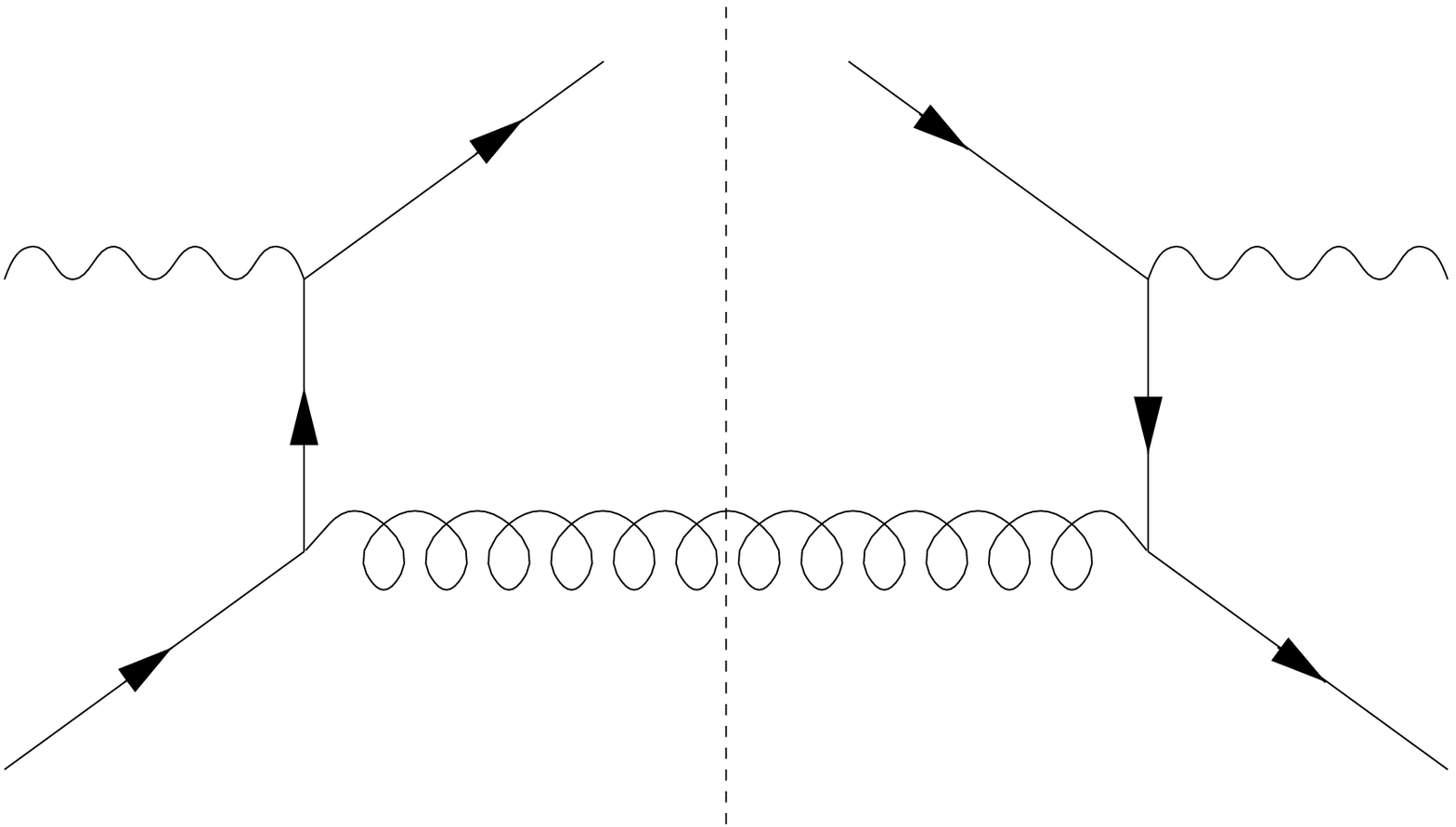} \hspace{2cm}
\includegraphics[width=4cm]{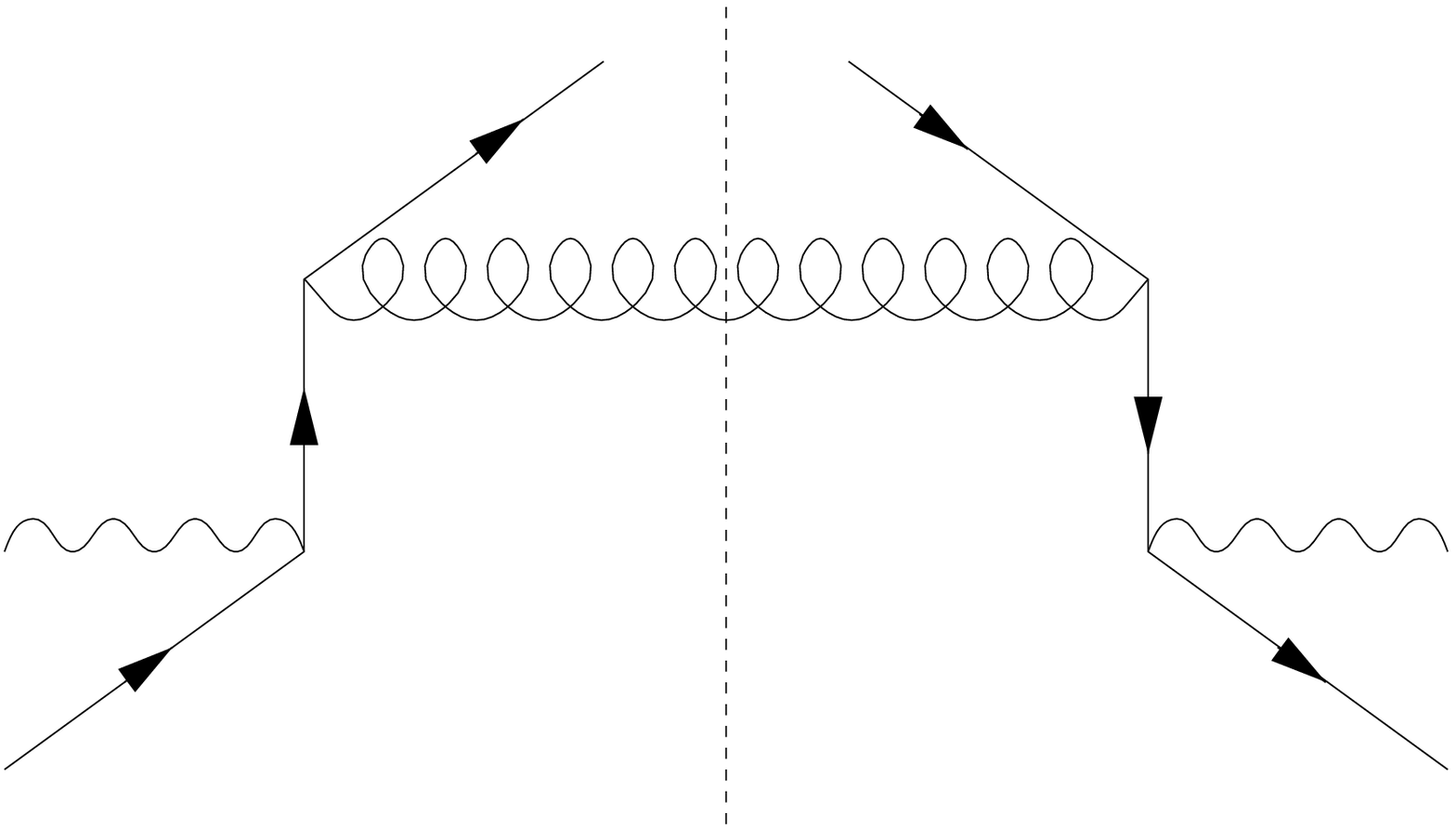}
\end{center}
\caption{Amplitudes which contribute to the SIDIS cross section
in which a hard gluon is radiated.\label{sidisGluon}}
\end{figure}

The first example that we will give  here is SIDIS at large $p_T$,
for which the leading hard parts are given in Fig.~\ref{sidisGluon}.
One can distinguish this process from ordinary SIDIS by looking at two jet 
production. 
The longitudinal gluons coming from the incoming proton are collected into links. 
These gluons couple to the radiated gluon, 
the fragmenting quark and the off-shell quark in the hard process.
Since there are no incoming lines to which the longitudinal gluon can couple, 
loops will not appear. 
A calculation taking all possible insertions into
account confirms this and
shows that the link structure is similar to that in ordinary SIDIS
\begin{equation}
\mbox{large $p_\st$ SIDIS}:\quad 
\frac{T_F\,N_c}{C_F}\,\mathcal U^{[+]}\,
\frac{1}{N_c}\tr^C\bigl[\mathcal U^{[+]\dagger}\mathcal U^{[+]}\bigr]
- \frac{T_F}{C_F\,N_c}\,\mathcal U^{[+]} = \mathcal U^{[+]}\ ,
\end{equation}
where the symbol $\tr^C$ is used to distinguish color tracing from Dirac tracing and $T_F$ = 1/2 and $C_F$ = $T_F (N_c^2 -1)/N_c$ = 4/3 in $SU(3)$.
The weighted transverse moment appearing in an appropriate azimuthal
asymmetry is therefore $\Phi_\partial^{[+]\,\alpha}$.

The second process considered here is DY at large $p_\st$, 
where the situation is more complicated. 
In simple DY there is only one (incoming) line to which longitudinal gluons can couple, 
which leads to a $\mathcal U^{[-]}$--link. 
\begin{figure}
\begin{center}
\includegraphics[width=4cm]{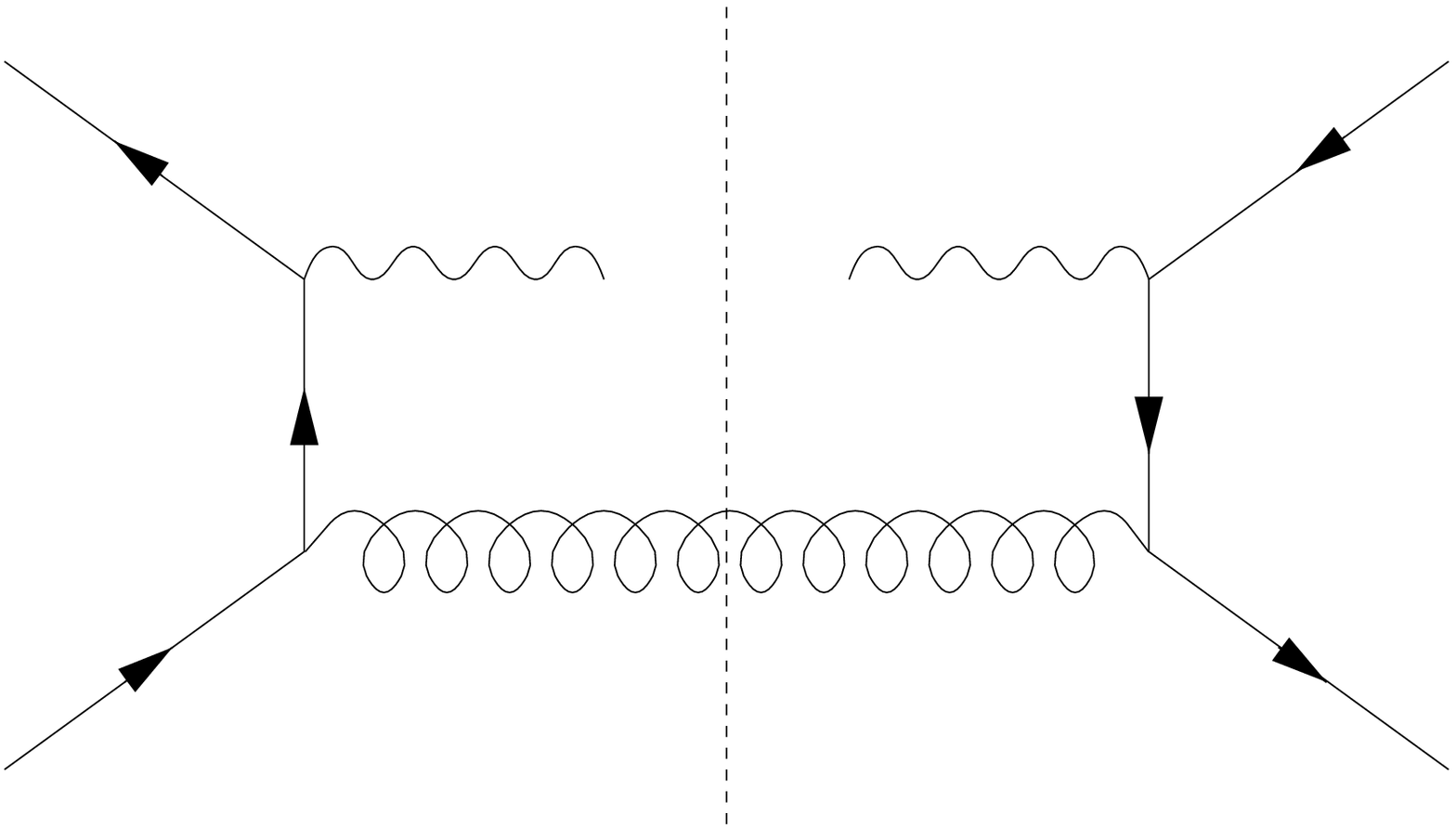} \hspace{2cm}
\includegraphics[width=4cm]{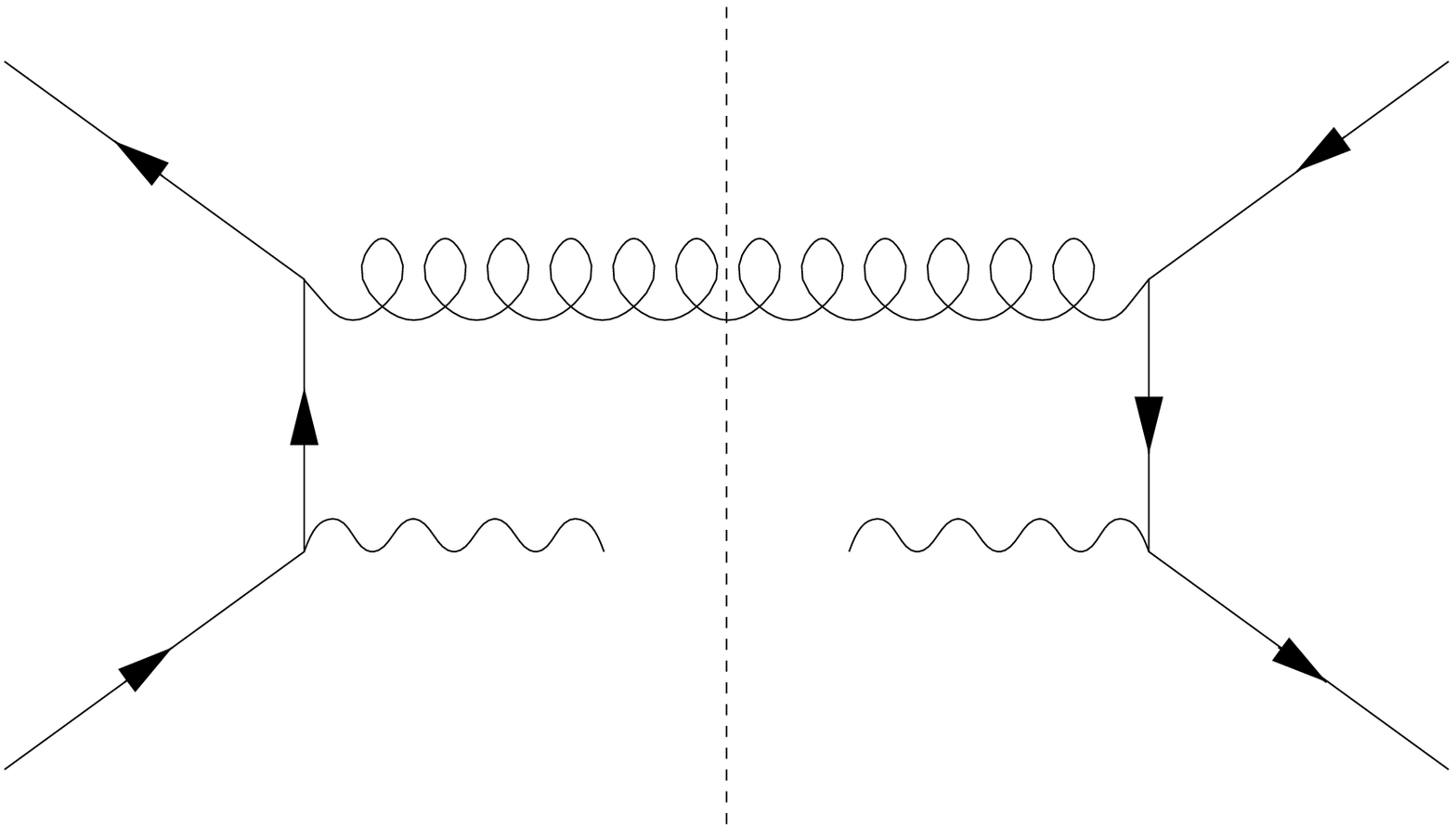}
\end{center}
\caption{Amplitudes which contribute to the DY cross section
in which a hard gluon is radiated.\label{DYGluon}}
\end{figure}
However, if a hard gluon is radiated (see Fig.~\ref{DYGluon}),
the link structure changes completely. 
The outgoing line together with the incoming line give in general  the possible presence of a loop and this is exactly what occurs here. 
The calculation leads to a link structure in the quark-quark correlator for the
lower hadron of the form
\begin{equation}
\mbox{large $p_\st$ DY}:\quad 
\frac{T_F\,N_c}{C_F}\,\mathcal U^{[+]}
\,\frac{1}{N_c}\tr^C\bigl[\mathcal U^{[\Box]\dagger}\bigr]
- \frac{T_F}{C_F\,N_c}\,\mathcal U^{[-]} 
= \frac{9}{8}\,\mathcal U^{[+]}
\,\frac{1}{N_c}\mathrm{Tr^C}\bigl[\mathcal U^{[\Box]\dagger}\bigr]
- \frac{1}{8}\,\mathcal U^{[-]}\ .
\end{equation}
The explicit links and their differences only appear in the 
(unintegrated) transverse momentum dependent parton distributions. 
After integration over transverse momenta, 
the differences vanish in $\Phi(x)$.
Weighing with $p_\st$ once, 
one obtains the transverse moments $\Phi_\partial^\alpha(x)$ and, 
as we have seen in Eq.~\eqref{glp}, 
these differences may lead to an enhancement of the gluonic pole contribution in the transverse moment.
In QCD, however, this is not the case for traced loops.
For a loop in the gauge link $\mathrm{Tr^C}\big[\mathcal U^{[\Box]}\big]$, 
one obtains a structure in the integrand of the form
$\mathrm{Tr^C}\big[\mathcal U\,G^{+\alpha}\,\mathcal U\big]$, which vanishes.
Therefore the transverse moment appearing in large $p_\st$ DY has
the link structure
$\frac{9}{8}\,\Phi_\partial^{[+]\,\alpha}
-\frac{1}{8}\,\Phi_\partial^{[-]\,\alpha}$.
It differs from the ordinary DY process, 
where the transverse moment is $\Phi_\partial^{[-]\alpha}$. 

We mention here two other examples, namely processes with 
quark-antiquark and quark-quark scattering as hard processes,
relevant in processes like pion production in hadron-hadron scattering. 
Each of the soft parts in such a process gets a particular link structure.
Considering for hadron-hadron scattering only the contribution
coming from quark-antiquark scattering given in Fig.~\ref{fig-qqbarlinks}a,
with in the t-channel now gluon exchange, one finds that
all four link structures in Fig.~\ref{fig-qqbarlinks} are
involved due to the color-flow,
\begin{equation}
\mbox{Fig.~\ref{fig-qqbarlinks}a in QCD}:\quad 
\frac{1}{8}\,\mathcal U^{[+]}
\,\frac{1}{N_c}\mathrm{Tr^C}\bigl[\mathcal U^{[\Box]\dagger}\bigr]
+ \frac{7}{8}\,\mathcal U^{[-]}.
\end{equation}
For quark-quark scattering one finds
\begin{equation}
\mbox{Fig.~\ref{fig-qqlinks}a in QCD}:\quad 
\frac{5}{4}
\frac{1}{N_c}\mathrm{Tr^C}\bigl[\mathcal U^{[\Box]\dagger}\bigr]
\,\mathcal U^{[+]}
- \frac{1}{4}\,\mathcal U^{[\Box]}\,\mathcal U^{[+]},
\end{equation}
where the latter term is an example of an open loop, which,
in analogy to Eq.~\eqref{glp},
gives an enhancement of the 
gluonic pole contribution as compared to $\mathcal U^{[+]}$.

In the above examples we have only been concerned with the link 
structure of the correlator $\Phi$, which describes
the embedding of
an incoming quark. The link structure arose from longitudinal
collinear gluons. In the same way the other correlators must be
dealt with and will acquire particular link structures, which
also may involve loops. One of those correlators, 
which describes the fragmentation of a quark in SIDIS at large $p_\st$, 
receives for instance a loop contribution. Since there are two sources
for T-odd effects on the fragmentation side, the transverse moment as given
in Eqs.~\eqref{glp0} and \eqref{glp} consists in this case
of two terms, one coming from
gluonic poles and one from final state interactions. 
As argued in~\cite{Boer:2003cm}, 
the combination of these two mechanisms spoils
the simple sign relation between unintegrated fragmentation functions
appearing in SIDIS and electron-positron annihilation
in Refs~\cite{Metz:2002qg, Brodsky:2002ue}.

\section{Conclusions}

In this paper we have shown that loops can appear in link structures in 
the correlation functions used in hard 
processes. These link structures become relevant as soon as one is
sensitive to the transverse momenta of the partons. 
The loops may contribute to the transverse moments as enhanced gluonic poles.
For distribution functions, gluonic poles are the origin of the T-odd distribution functions appearing in single spin asymmetries.
In order to show the appearance of loops we started 
with an abelian version of QCD. However, also in non-abelian theories, 
loops appear for which a few explicit results have been given. 

The link structures can be derived by resumming all longitudinally 
polarized gluons, coming from a soft blob, into a gauge invariant 
correlation function.
Although this seems to break universality, 
this needs not to be the case in $p_\st$-integrated and $p_\st$-weighted quantities, 
which lead to well-defined bilocal lightcone matrix elements.
Nevertheless,
in the study of factorization (see e.g.~\cite{Ji:2004wu, Ji:2004xq})
and the evolution of transverse momentum
dependent distribution and fragmentation functions~\cite{Henneman:2001ev}
one also needs to
account for the structure and appearance of loops in the gauge links.
In principle things look fine for SIDIS, 
because we found that one is always dealing with a $\mathcal U^{[+]}$-link, but in DY 
a gluon ladder leads to different link structure. 
Also for fragmentation in SIDIS one finds such differences.

We have discussed the link structures appearing in large $p_\st$ SIDIS and
large $p_\st$ DY processes. 
We note that having found the link structures, there is still work to
be done in finding out the specific 
observables in which one sees their effects. One
needs an observable sensitive to the intrinsic
transverse momentum of the partons, which is different from the large 
$p_\st$~\cite{Boer:2003tx}.
T-odd observables, such as the L-R asymmetry in $p p^\uparrow \rightarrow
\pi X$~\cite{Anselmino:1995tv, Anselmino:1998yz, 
Anselmino:2002pd, Boglione:1999pz}, may play an important role here, 
as they are absent for integrated distribution functions at leading order.

\section{Acknowledgements}

We acknowledge discussions with D.~Boer, A.~Sch\"afer and W.~Vogelsang. Part
of this work
was supported by the foundation for Fundamental Research of Matter (FOM) and
the National Organization for Scientific Research (NWO).

\bibliographystyle{apsrev}
\bibliography{references}

\end{document}